\documentclass{ieeeconf}

\usepackage{authblk}
\usepackage{cite}
\usepackage{amsmath}
\usepackage{epsf,latexsym,amssymb}
\usepackage{graphicx}
\usepackage{amsmath}  %will give errors with \be \ee !!!
\usepackage{color}
\usepackage[margin=1in]{geometry}

\newcommand{\R}{{\mathbb R}}  %ams bold

\newcommand{\be}[1]{\begin{equation}\label{#1}}
\newcommand{\ee}{\end{equation}}

\newcommand{\mybox}{\hfill $\Box$} %put \qed at right margin (white square)
\newcommand{\beqn}{\begin{eqnarray*}}
\newcommand{\eeqn}{\end{eqnarray*}}
\newcommand{\beqnum}{\begin{eqnarray}}
\newcommand{\eeqnum}{\end{eqnarray}}
\newcommand{\ben}{\begin{enumerate}}
\newcommand{\een}{\end{enumerate}}

\newcommand{\xnonlin}{\xi}
\newcommand{\xlin}{\tilde\xi } % state, to be redefined perhaps
\newcommand{\x}{\xi } 

\newtheorem{theorem}{Theorem}
\newtheorem{itlemma}{Lemma}[section] %number by section (set in \em by default)
\newtheorem{itproposition}[itlemma]{Proposition}
\newtheorem{itcorollary}[itlemma]{Corollary}
\newtheorem{itremark}[itlemma]{Remark}
\newtheorem{itdefinition}[itlemma]{Definition}

\newenvironment{lemma}{\begin{itlemma}\rm}{\end{itlemma}} %no-italics
\newenvironment{remark}{\begin{itremark}\rm}{\end{itremark}} %no-italics
\newenvironment{corollary}{\begin{itcorollary}\rm}{\end{itcorollary}}
\newenvironment{proposition}{\begin{itproposition}\rm}{\end{itproposition}}
\newenvironment{definition}{\begin{itdefinition}\rm}{\end{itdefinition}}

\newcommand{\bl}[1]{\begin{lemma}\label{#1}}
\newcommand{\br}[1]{\begin{remark}\label{#1}}
\newcommand{\bt}[1]{\begin{theorem}\label{#1}}
\newcommand{\bd}[1]{\begin{definition}\label{#1}}
\newcommand{\bp}[1]{\begin{proposition}\label{#1}}
\newcommand{\bc}[1]{\begin{corollary}\label{#1}}
\newcommand{\ec}{\mybox\end{corollary}}
\newcommand{\el}{\end{lemma}}
\newcommand{\er}{\mybox\end{remark}}
\newcommand{\et}{\end{theorem}}
\newcommand{\ed}{\mybox\end{definition}}
\newcommand{\ep}{\mybox\end{proposition}}
\newcommand{\epr}{\end{proof}}
\newcommand{\bpr}{\begin{proof}}

\newcommand{\fref}[1]{Fig.~\ref{#1}}

\def\twodigits#1{\ifnum#1<10 0\fi\the#1}
\makeatletter
  \newcommand*\short[1]{\expandafter\@gobbletwo\number\numexpr#1\relax}
\makeatother
\usepackage{datetime}

\usepackage[colorlinks=true, allcolors=blue]{hyperref}

\title{An exact active sensing strategy for a class of bio-inspired systems}

\author{Debojyoti Biswas$^{1}$, Eduardo D. Sontag$^{2,*}$, Noah J. Cowan$^{1,3,*}$
\thanks{$^{1}$Laboratory for Computational Sensing and Robotics, Johns Hopkins University, Baltimore, Maryland, United States. {\tt\small dbiswas2@jhu.edu}}%
\thanks{$^{2}$Department of Electrical and Computer Engineering and Department of Bioengineering, Northeastern University, Boston, Massachusetts, United States. {\tt\small e.sontag@northeastern.edu}}
\thanks{$^{3}$Department of Mechanical Engineering, Johns Hopkins University, Baltimore, Maryland, United States. {\tt\small ncowan@jhu.edu}}%
\thanks{$^{*}$Contributed and co-supervised equally.}
}

\begin{document}
\maketitle

\medskip
\begin{abstract}
We consider a general class of translation-invariant systems with a specific category of output nonlinearities motivated by biological sensing. 
We show that no dynamic output feedback can stabilize this class of systems to an isolated equilibrium point. To overcome this fundamental limitation, we propose a simple control scheme that includes a low-amplitude periodic forcing function akin to so-called ``active sensing'' in biology, together with nonlinear output feedback.
% We demonstrate that this approach leads to the emergence of an exponentially stable limit cycle. These analyses provide a provably stable active sensing scheme and may thus help rationalize active sensing movements made by animals as they perform certain motor behaviors.
Our analysis shows that this approach leads to the emergence of an exponentially stable limit cycle. These findings offer a provably stable active sensing strategy and may thus help
to rationalize the active sensing movements made by animals as they perform certain motor behaviors.
\end{abstract}
%\begin{IEEEkeywords} ... \end{IEEEkeywords}

\section{Introduction}

Biological sensory systems often 
exhibit an attenuated response to 
constant (DC) stimuli, allowing such biosensors to excel at detecting changes (AC) rather than measuring absolute values \cite{taylor2007sensory}.
This feature, often referred to as sensory adaptation, poses significant challenges for conventional state estimation and control. For specific examples related to control theory and systems biology, see  \cite{sontag2003adaptation,sontag2008remarks}. To overcome this sensory adaptation, animals appear to use ancillary movements, referred to as \textit{active sensing} movements, that drive robust responses in their change-detecting sensory systems \cite{gibson1962observations,bajcsyactive1988,schroeder2010dynamics}. Animals use this strategy to enhance sensory information across sensory modalities, e.g.,\ echolocation \cite{wohlgemuth2016action}, whisking \cite{mitchinson2011active} and other forms of touch \cite{prescott2011active,saig2012motor}, electrosense \cite{hofmann2013sensory,chen2020tuning,stamper2012active}, and vision \cite{ahissar2012seeing,michaiel2020dynamics}. It is well established that conditions of decreased sensory acuity lead to increased active movements \cite{lockey2015one, stockl2017comparative, deora2021tactile, stamper2012active, chen2020tuning, michaiel2020dynamics, wohlgemuth2016action, kiemel2002multisensory, rucci2015unsteady}, suggesting a closed-loop perceptual process \cite{biswas2018closed,ahissar2016perception}.

The ubiquity of active sensing in nature motivated us to explore the mathematical conditions that might necessitate active sensing. One theory is that active sensing is at least in part borne out of the need for nonlinear state estimation \cite{biswas2023mode, sontag2022observability}. Under this theory, animals use active sensing---that is, the generation of time-varying motor commands that continuously stimulate their sensory receptors---so that the system states can be estimated from sensor measurements. A complementary approach---and one we pursue in this paper---is that active movements do indeed enhance observability, but that full state estimation itself may be unnecessary. In other words, active sensing movements may enable stabilizing output feedback without recourse to state estimation as an intermediate step.

In this paper, we examine a class of systems with a nonlinear sensory output that mimics sensory adaptation and perceptual fading in nature \cite{taylor2007sensory,fabrelarge2020,riggs1953disappearance} resulting in a system whose linearized dynamics is unobservable \cite{kunapareddy2018recovering}  (Section \ref{sec:bio}).  In essence, the usual state-estimate-based control framework that dominates engineering practice in many fields \cite{barfoot2024state} cannot be na\"ively applied. More fundamentally,  we show that the class of bio-inspired nonlinear models considered here cannot be stabilized around an equilibrium point with any choice of dynamic output feedback (Section \ref{sec:impossibility}). However, with appropriate control inputs, nonlinear observability can persist, allowing us to mimic active sensing behavior observed in animals \cite{biswas2018closed,biswas2023mode}. Specifically,  we present an active-sensing-based output feedback system (Section~\ref{sec:exact}), prove that it stabilizes an arbitrarily small limit cycle (Section \ref{sec:lyap}), and numerically characterize the nonlinear system dynamics (Section~\ref{sec:numerical}).

\begin{figure}[b]
  \centering
  \includegraphics[width=\columnwidth]{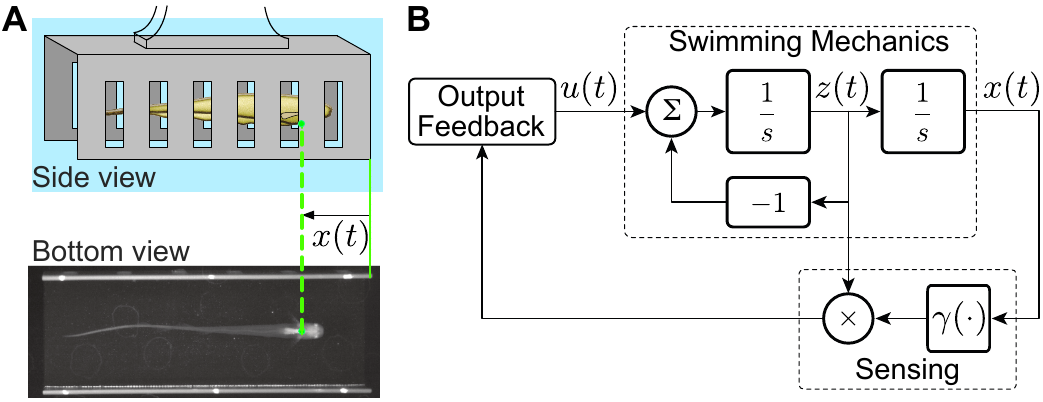}
  \caption{(A)  Weakly electric fish control their position using active sensing to remain within a refuge; $x(t)$ is the fish's position relative to the refuge. (B) Simplified model.}
  \label{fig:eigen}
\end{figure}

\section{Biologically inspired system definition} \label{sec:bio}
Station keeping behavior in weakly electric fish, \textit{Eigenmannia virescens}, provides an ideal system for investigating the interplay between active sensing and task-level control \cite{stamper2012active,biswas2018closed,chen2020tuning,yeh2024illumination}; see \fref{fig:eigen}. These fish routinely maintain their position relative to a moving refuge and use both vision and electrosense to collect the necessary sensory information from their environment \cite{cowan2007critical, cowan2014feedback,sutton2016dynamic,roselongitudinal1993}. While tracking the refuge position (i.e., task-level control), the fish additionally produce rapid ``whisking-like'' forward and backward swimming movements (i.e., active sensing). When vision is limited (for example, in darkness), the fish increase their active sensing movements \cite{stamper2012active,biswas2018closed,yeh2024illumination}, likely to excite their change-detecting, high-pass electroreceptors \cite{nelson1997characterization}.

To capture the essence of this behavior, suppose $x$ is the position of the animal and $z = \dot x$ is its velocity as it moves in one degree of freedom. We assume that a sensory receptor measures only the local rate of change of a stimulus, $s(x)$ as the animal moves relative to the sensory scene, i.e., $y=\frac{d}{dt}s(x)$. Defining $\gamma(x) := \frac{d}{dx}s(x)$, we arrive at a $2$-dimensional, single-input, single-output normalized mass-damper system of the following form \cite{sefati2013mutually,kunapareddy2018recovering}:
\begin{equation}
\label{eq:simple_system}
\begin{aligned}
\dot x &= z, \quad && x\in \R\\
\dot z &= -z \,+\, u, \quad && z,u\in \R\\
y   &= \frac{d}{dt}s(x)=\gamma(x)\,z, \quad && y\in \R
\end{aligned}
\end{equation}
where the mass and the damping constant are both assumed to be unity. Linearization of the above system (\ref{eq:simple_system})  around any equilibrium, $(x^*, 0)$, is given as follows:

\begin{align*}
\dot{\xlin} &= \underbrace{\begin{bmatrix}
0 &1\\
0 &-1
\end{bmatrix}}_{:=A}\xlin+ \underbrace{\begin{bmatrix}
0 \\
1
\end{bmatrix}}_{:=B}u\\ 
y& =  \underbrace{\begin{bmatrix}
0 & \gamma^*
\end{bmatrix}}_{:=C}\xlin,
\end{align*}
where
$\xlin = (x-x^*,\;z)^{\top}$ and $\gamma^* = \gamma(x^*)$. Clearly (A,C) is not observable irrespective
of $\gamma^*$\cite{rugh1996linear}.
Indeed, the output introduces a zero at the origin that
cancels a pole at the origin, rendering $x$ unobservable. 
Assuming no input $u$, we can write the system (\ref{eq:simple_system}) as, \be{eq:simple-nonlin} \dot \xnonlin = f(\xnonlin),
\quad y = h(\xnonlin), \ee where
$\xnonlin = (x,\;z)^{\top},\;f = (z,\;-z)^{\top}$ and
$h(\xnonlin) = \gamma(x)z$. We can construct the observation space,
$\mathcal{O}$ (set of all infinitesimal observables) by taking
$y = \gamma(x)z$ with all repeated time derivatives
\[
y^{(k)} = L_f^{(k)}(\gamma(x)z)
\]
as in \cite{mct,nijmeijer1990nonlinear}.  The
superscript ``$(k)$'' indicates $k$th order derivative. Note that
$L_f^{(k)}((\gamma(x)z))$ lies in the span of the functions
$\gamma^{(j)}(x)z^{j+1},\,j = 0,1,\ldots, k$. The rank condition on the
observability co-distribution \cite{mct,nijmeijer1990nonlinear} implies a sufficient condition for local
observability as follows
\cite{kunapareddy2018recovering}:
\be{}
    z^2(2(\gamma^{\prime}(x))^2-\gamma(x)\gamma^{\prime\prime}(x))\neq 0.
\ee 
For an non-hyperbolic $\gamma$ (i.e.\ $\gamma\neq1/(c_1 x+c_0)$, with
constants $c_0,c_1$), the non-zero velocity
requirement, $z\neq 0$, implies the need for active sensing to maintain the local
observability of the system~\cite{kunapareddy2018recovering}.
%The condition for global nonlinear observability further requires the aperiodicity of $\gamma$ \cite{sontag2022observability}.
For the condition for global nonlinear observability see \cite{sontag2022observability}.

Given that, under the conditions described above, the system is
 locally nonlinearly observable, can we design a nonlinear output
feedback controller, that can stabilize the system to an equilibrium point? The next section addresses this question.

\section{An impossibility result for stabilizing the system to a point}
For the system in~\eqref{eq:simple_system}, dynamic output feedback cannot asymptotically
stabilize the origin $(0,0)$, as shown in the following proposition.

\begin{figure}[htbp!]
    \centering
    \includegraphics[width= 0.5\columnwidth]{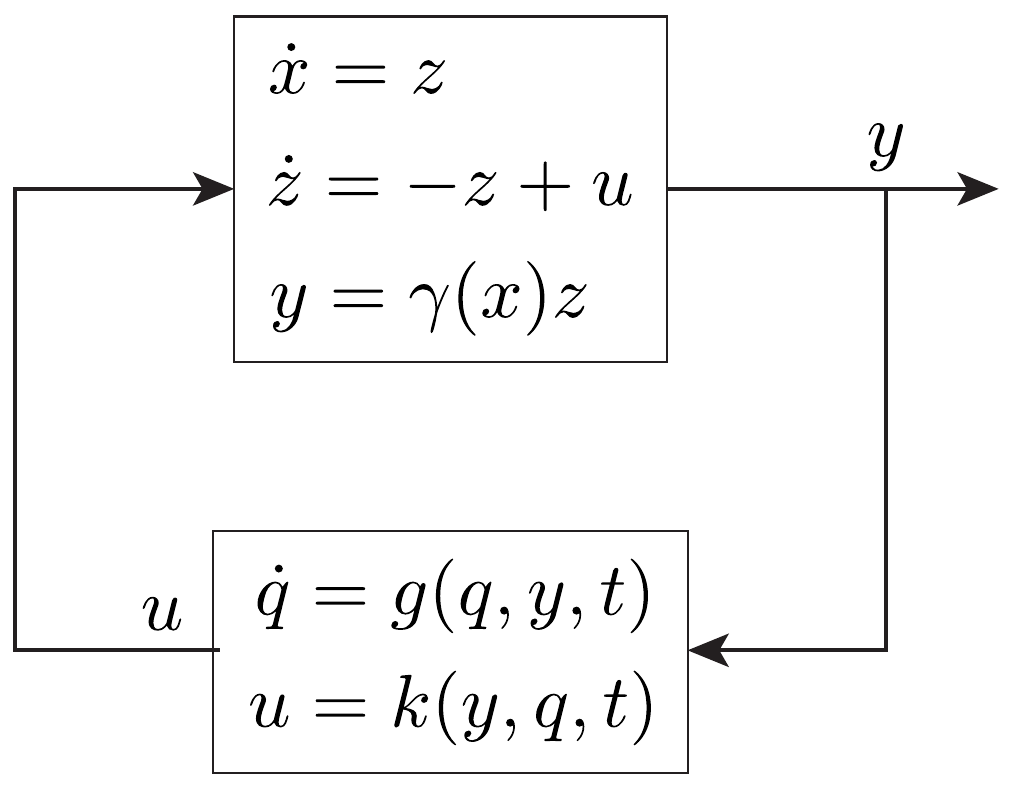}
    \caption{The system (\ref{eq:simple_system}) cannot be stabilized to an equilibrium
      point by the dynamic feedback in (\ref{eq:outputfeedback}).}
    \label{impossible}
\end{figure}

\bp{prop:nooutputfeedback}\label{sec:impossibility}
Consider the system 
(\ref{eq:simple_system}). Let
\begin{equation}
  \label{eq:outputfeedback}
  \begin{aligned}
  \dot q &= g(q,y,t)\\
  u&=k(y,q,t)
\end{aligned}
\end{equation}
be a dynamic, potentially time-varying, output feedback (\fref{impossible}). 
Suppose $(x^*,z^*,q^*(t)))=(0,0,q^*(t))$ is a solution to the
coupled system. Then there is a continuum of solutions,  $(\xi^*,0,q^*(t))$, $\xi^*\in\mathbb{R}$.
\ep

\bpr Since $(0,0,q^*(t))$ is a solution, we see
from the second equation in (\ref{eq:simple_system}) that
$u^*(t)=k(0,q^*(t),t)\equiv0$.  That means that $k(\gamma(\xi^*)\cdot 0,q^*(t),t)=0$, i.e.\
$(\xi^*,0,q^*(t))$ is also a solution, for all $\xi^*\in\mathbb{R}$.
\epr

In other words, no matter how ``fancy'' one makes the output feedback, there will always be a continuum of equilibria, and thus stabilizing one's favorite equilibrium among them is impossible.  In the next section, we achieve the next best thing: by adding a time-varying   \textit{active sensing} input to an output feedback term, we stabilize an arbitrarily small limit cycle.

\section{An exact active sensing strategy} \label{sec:exact} For $z\neq0$, the system (\ref{eq:simple_system}) is locally observable but cannot be stabilized to a point using output feedback, raising the question of whether it is possible to create a ``small" stable periodic orbit as the ``next best thing" to stabilizing a point.  To explore this, we make two simplifying assumptions:
\begin{enumerate}
\item[(A1)] The position-dependent scene is locally quadratic, namely $s(x) = \frac{1}{2}x^2$, leading to $\gamma(x) = x$.
\item[(A2)] The velocity, $z$, is directly measurable.
\end{enumerate}
These assumptions lead to the following simplified system with an augmented output equation
\begin{equation}
\label{eq:aug_system}
\begin{aligned}
\dot x &= z, \quad && x\in \R\\
\dot z &= -z \,+\, u, \quad && z,u\in \R\\
  y &   =
\begin{bmatrix}
  y_1 \\ y_2
\end{bmatrix}=
\frac{d}{dt}
\begin{bmatrix}
s(x)  \\ x
\end{bmatrix}=
\begin{bmatrix} xz \\ z
\end{bmatrix}, \quad && y\in\R^2.
\end{aligned}
\end{equation}

Note that the simplified system with augmented output \eqref{eq:aug_system} remains linearly unobservable, and following the same reasoning from Proposition \ref{sec:impossibility}, this system is also not stabilizable via any dynamic output feedback.
Thus adding velocity sensing and simplifying the measurement nonlinearity \emph{do not mitigate the lack of observability nor enable output stabilization}, which were the main challenges associated with the bio-inspired sensor. However, the assumptions simplify the following exposition.

\begin{figure}[htbp!]
\centering
\includegraphics[width = \columnwidth]{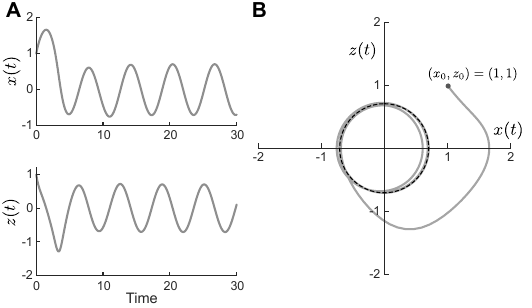}
\caption{Evolution of the system states, $x(t)$ and $z(t)$ for $\delta = 1/2$ with $k =1, a= 1/\sqrt{2}$ from initial condition $(x_0,z_0)=(1,1)$. (A) Time traces. (B) State trajectories on $x$-$z$ plane. The black dashed line represents the steady-state circular orbit of radius $a$.
}
\label{fig:numsoln}
\end{figure}

In the system \eqref{eq:aug_system}, it is easily verified that setting the input $u(t)$ to $\alpha(t) = a \cos(t)- a\sin(t)$ leads to a family of periodic solutions of the form $x(t) = a \sin(t) + C$ and $z(t)=a\cos(t)$, $C\in\R$. Our goal is to incorporate an output feedback term that stabilizes the system to the solution with $C=0$. With this in mind, consider the following ``active-sensing'' based controller:

\begin{equation}\label{forced_simple_system1}
u(t)=  \overbrace{~~\alpha(t)~~}^{\text{active~sensing}} -\quad \overbrace{k\big(F(y)-F(y^*)\big)}^{\text{output feedback}},
\end{equation}
where $F(y) = y_1y_2$ and $F(y^*) = y_1^*(t)y_2^*(t)$. Here, the active sensing input $\alpha(t)$ is a feed-forward term that maintains observability \cite{2007uniform}.
This leads to a single periodic solution
\[
  (x^*(t),z^*(t))=(a\sin(t),a\cos(t))
\]
with associated periodic output
$y^*_1(t)=x^*(t)z^*(t)$ and $y^*_2(t)=z^*(t)$. As we will show, the feedback term
$k(F(y)-F(y^*)) = k(y_1y_2-y_1^*(t)y_2^*(t))$ ensures the system converges to $\x^*(t)=(x^*(t),z^*(t))$ for appropriate choices of $a$ and $k$. 
We can rewrite the system \eqref{eq:aug_system} with input \eqref{forced_simple_system1} as follows: 
\begin{equation}\label{eq:forced_simple_system}
\begin{aligned}
 \dot{x}=&\, z \\   
  \dot{z}  =&\, -z - k(xz^2 - a^3\sin(t)\cos^2(t))\\
  &\, +a\cos(t)-a\sin(t)
\end{aligned}
\end{equation}
A numerical example showing the  system's states converge to a circular orbit of
radius $a$ is shown in (\fref{fig:numsoln}). 

Linearization of 
(\ref{eq:forced_simple_system}) around $\x^*(t)$ results in a
linear $\pi$-periodic system:
\begin{gather}
\label{eq:LTV}
  \dot \xlin = A(t)\xlin,
  \text{ where }~\xlin:=\xnonlin-\x^*\\ %= (x-x^*,z-z^*)^\top,\\
  A(t):=\begin{bmatrix}
    0 &1\\
    -\delta\cos^2(t) &-1-\delta\sin(2t)\\
  \end{bmatrix},
  \intertext{with the parameter $\delta$ is defined as}
  \delta:=ka^2.
\end{gather}
The linear $\pi$-periodic system (\ref{eq:LTV}) is parameterized by $\delta = ka^2$, which depends on both the choice of the output feedback gain, $k$, and square of the radius of the circular active sensing orbit, $a$. In the following sections, we analyze the stability of the system (\ref{eq:LTV}) using Lyapunov and Floquet theory.

\section{Lyapunov stability of linearized, time-periodic  system}
\label{sec:lyap}
\bt{theo:SISL}
The origin of the system \eqref{eq:LTV}  is stable in the sense of Lyapunov for $0<\delta\leq  \frac{2}{3} (-2 + \sqrt{7})=:\delta^\dagger$.
\et

\bpr
We consider a quadratic Lyapunov candidate function, $V(\xlin(t))$ given by
\begin{equation}
  \label{eq:lyap}
\!\!V(\xlin(t)) = \dfrac{1}{2}\xlin(t)^\top P \xlin(t),\,
\text{}\, P = \begin{bmatrix}
    1 &1 \\
   1 &\eta
\end{bmatrix},\, \eta>1.
\end{equation}
The derivative of $V$ along the trajectories of the linear system (\ref{eq:LTV}) is given by
\begin{align}%\label{eq:lyapderiv}
    \dot V(\xlin(t)) %&= \dfrac{1}{2}\xlin(t)^\top(PA(t)+A(t)^\top P)\xlin(t)\\
                  = -\xlin(t)^\top Q(t)\xlin(t),
\end{align}
where $Q(t):=-\frac{1}{2}(PA(t)+A(t)^\top P)$ is
\begin{equation*}
  \begin{bmatrix}
    \delta\cos^2(t) & \frac{\delta}{2} (\sin(2t) + \eta\cos^2(t)) \\
    \frac{\delta}{2} (\sin(2t) + \eta\cos^2(t)) &  \delta\eta\sin(2t) + (\eta-1)
  \end{bmatrix}.
\end{equation*}
\iffalse
\begin{multline*}
  \!\!\!Q(t) =\\
  \begin{bmatrix}
    \delta\cos^2(t) & \frac{\delta}{2} (\sin(2t) + \eta\cos^2(t)) \\
    \frac{\delta}{2} (\sin(2t) + \eta\cos^2(t)) &  \delta\eta\sin(2t) + (\eta-1)
  \end{bmatrix}.
\end{multline*}
\fi

Note that $Q(t)$ is positive semidefinite (a sufficient condition for Lyapunov stability) if the trace and determinant are nonnegative. Starting with the trace and assuming $\delta>0$:
\begin{equation*}
  \mathrm{Tr}(Q) = \delta\cos^2(t) + (\eta-1) + \eta\delta\sin(2t)\geq (\eta-1)-\eta\delta.
\end{equation*}
Thus if $\delta<(\eta-1)/\eta$ then $\mathrm{Tr}(Q)>0$.  The determinant is given by
\begin{equation*}
\begin{aligned}
\mathrm{Det}(Q) =&\, -\frac{1}{4} \delta\cos ^2(t)(4(1-\eta)\\
&\,+\eta ^2 \delta \cos ^2(t)-2\eta\delta\sin(2t)+4 \delta\sin^2(t)).    
\end{aligned}
\end{equation*}
Since $-(1/4)\delta\cos^2(t)$ is negative $\forall t$ except where it vanishes at $k\pi$, $k\in\mathbb{Z}$, we focus on ensuring that the second term is also negative. If we assume $\delta< 4(\eta-1)/(\eta^2+2\eta+4)$ then
\begin{align*}
& 4(1-\eta) +\eta ^2 \delta \cos ^2(t)-2\eta\delta\sin(2t)+4 \delta\sin^2(t)\\
&<4(1-\eta) +\delta (\eta^2+2\eta+4 )\\
 &<0
\end{align*}
ensures  $\mathrm{Det}(Q)\ge 0$ (only vanishing at $t= k \pi$,
$k\in\mathbb{Z}$).
It is easy to show that this constraint  $\delta\leq 4(\eta-1)/(\eta^2+2\eta+4)$  implies the constraint from the analysis of the trace, namely $\delta<(\eta-1)/\eta$. Note that  $\max_{\eta>1}4(\eta-1)/(\eta^2+2\eta+4) = \frac{2}{3} (-2 + \sqrt{7})=\delta^\dagger$ occurs at $\eta^\dagger=1+\sqrt{7}$. Thus, choosing $\eta=\eta^\dagger$ in our candidate Lyapunov function, then if  $0<\delta\leq \delta^\dagger$, the $Q$ matrix is positive definite $\forall t$ except where it becomes semi-definite at one instant per period. Hence $\dot V\leq 0$ and the proof is complete.
\epr

\begin{remark}
  The proof for Theorem \ref{theo:SISL} implies whenever $\xlin\neq 0$
  then the Lyapunov function is strictly decreasing, i.e.,\
  $\dot V<0$, except at one instant per period 
  where $\dot V =0$. This will be useful in the following corollary.
\end{remark}

\bc{corollary:asymp_stable}
The origin of \eqref{eq:LTV} is asymptotically stable for $\delta\leq \delta^\dagger$
\ec

\bpr Consider the Lyapunov function from \eqref{eq:lyap} with
$\eta=\eta^\dagger$. For $\delta\leq\delta^\dagger$,
$\dot V\leq 0$ hence $V$ is a non-increasing function of time. Since
$V$ is also lower bounded by 0,
$\lim_{t\to \infty}V(t)=V_{\infty}\geq 0$. Consequently, since $V$ is
a positive definite, nonincreasing function of in $\xlin$ , $\xlin(t)$ is
bounded $\forall t\ge0$.

Note that
$\Ddot{V}(\xlin(t))=\xlin(t)^\top R(t)\xlin(t)$ with $R(t):=-(Q(t)A(t)+A(t)^\top Q(t))$ is bounded since $\xlin(t)$ is bounded and, by a
straightforward calculation, all the terms in $R(t)$ are bounded,
confirming that $\dot V$ is uniformly continuous in time. Hence by
Barbalat's lemma, $\lim_{t\to \infty}\dot V = 0$.

To argue that $\lim_{t\rightarrow\infty}\xlin(t)=0$ we will take a
point-wise approach.  Note that for a continuous function $f(t)$,
$\lim_{t\rightarrow\infty}f(t) =0$ $\iff$
$\lim_{k\rightarrow\infty}f(t_0+k\pi)=0$ $\forall t_0\in(0,\pi)$.
Also, note that since $Q(t)$ is periodic and positive definite
$\forall t$ except at $t=n\pi,n\in\mathbb{Z}$, it follows that
\[
  \lim_{k\rightarrow\infty}\xlin(t_0+k \pi)^\top
  Q(t_0)\xlin(t_0+k\pi)=\lim_{k\rightarrow\infty}\dot V(t_0+k \pi)= 0
\]
for all $t_0\in(0,\pi)$. Since for each $t_0\in(0,\pi)$ each fixed
matrix $Q(t_0)>0$, we have that
$\lim_{k\rightarrow\infty}\xlin(t_0+k\pi)=0$ for all $t_0\in(0,\pi)$. In
other words $\xlin(t)\rightarrow0$ as $t\rightarrow\infty$, which
completes the proof. \epr

\begin{remark} Exponential stability. Since \eqref{eq:LTV} is a continuous, linear, time-periodic system, its asymptotic stability implies the asymptotic stability of the corresponding time-invariant discrete-time map (i.e.,\
  the monodromy matrix). For a linear, time-invariant discrete-time system to be asymptotically stable, its eigenvalues ($\lambda_i$) must lie strictly within the unit circle and therefore the discrete-time system is also exponentially stable. The eigenvalues of the monodromy matrix are also called the
  Floquet multipliers of the system \cite{guckenheimer2013nonlinear}.
Thus, the corresponding Floquet exponents $\ln(\lambda_i)$ of the
  continuous time system will lie in the open left-half plane, implying
  that the continuous-time system is also exponentially stable. We will numerically compute the Floquet
multipliers for \eqref{eq:LTV} in the next section. 
\label{exp_stable}
\end{remark}

The Lyapunov stability analysis above has a few limitations. First, it is only local. Second, it leads to a somewhat conservative bound on $\delta$. Third, it does not address the convergence rate. Thus, we now turn toward numerical methods to examine local performance and characterize nonlinear stability.

%====================================================================
\section{Numerical stability analysis of active sensing controller} \label{sec:numerical}

\subsection{Linear stability analysis as a function of $\delta$}
\begin{figure*}[tb!] % Use figure* to span both columns
    \centering
    \begin{minipage}{0.64\textwidth} % Adjust the width as needed
        \centering
        \includegraphics[width=0.93\textwidth]{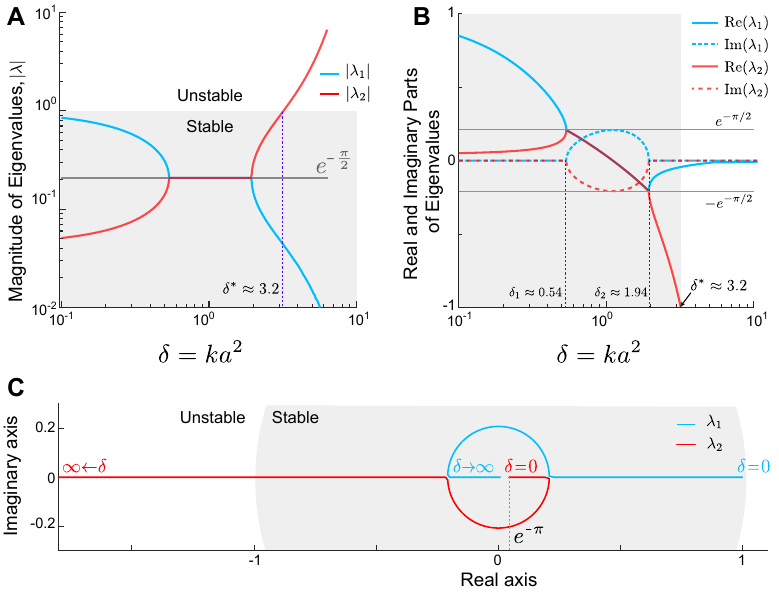}
    \end{minipage}%
    \hspace{1 em} % Horizontal space between figure and caption
    \begin{minipage}{0.32\textwidth} % Adjust the width as needed
        \vspace{-1 cm} % Aligns the caption to the top of the figure
        \caption{Eigenvalues of the linearized system (\ref{eq:LTV}) for different values of $\delta = ka^2$. (A) Modulus of the eigenvalues. The gray region represents the stable area, where the modulus of both eigenvalues is less than one. The gray solid line denotes the square root of the product of eigenvalues $(\exp(-\pi/2))$ and the blue dotted line denotes the critical value $\delta^* \approx 3.2$ above which the system (\ref{eq:LTV}) becomes unstable. (B) The real (solid line) and imaginary part (dashed line) of the eigenvalues. The eigenvalues are real for low values of $\delta$, become complex conjugates between $(\delta_1,\delta_2) = (0.54,1.94)$, and return to being real for higher values of $\delta$. (C) Evolution of the eigenvalues on the complex plane with increase in $\delta$. At $\delta = 0$, the eigenvalues are $1$ and $ \exp(-\pi)$, respectively. As $\delta \to \infty$ one eigenvalue approaches zero, while the other tends to infinity, with their product remaining constant $\exp(-\pi)$.  The gray regions in (B, C) are the same as in (A) representing the stable area.
}
\label{floquet}
    \end{minipage}
\end{figure*}

Suppose $\Phi(t)$ is the corresponding fundamental matrix of the system (\ref{eq:LTV}), constructed from the two linearly independent solution vectors $\begin{bmatrix}
    x_{11}(t) &z_{12}(t)
\end{bmatrix}^\top$ and $\begin{bmatrix}
    x_{21}(t) &z_{22}(t)
\end{bmatrix}^\top$ satisfying the initial conditions: %(to ensure $\Phi(0)=\mathrm{I}$)
\begin{align*}
 \begin{bmatrix}
    x_{11}(0)\\
    z_{12}(0)\\
    \end{bmatrix} = \begin{bmatrix}
    1\\
    0\\
    \end{bmatrix},\begin{bmatrix}
    x_{21}(0)\\
    z_{22}(0)\\
    \end{bmatrix} = \begin{bmatrix}
    0\\
    1\\
    \end{bmatrix}.
\end{align*}
Since $A(t)$ is of $\pi$-periodic, the monodromy matrix, $M$ is given by the evaluation of the fundamental solution matrix, $\Phi(t)$ at time $t =\pi$:
\begin{align*}
 M = \Phi(\pi)=\begin{bmatrix}
    x_{11}(\pi) &x_{21}(\pi)\\
    z_{12}(\pi) &z_{22}(\pi)
    \end{bmatrix} .
\end{align*}
The Wronskian, $W(t):=\det \Phi(t)$ satisfies $\dot{W}=\mathrm{tr}(A(t))W$. Hence integrating over 
$(0,\pi)$ we obtain $\det (M) = W(\pi) = e^{-\pi}$.
\iffalse
\begin{align*}
\int_0^\pi dW/W &= \int_0^\pi(-1-\delta\sin(2t))dt\\
\ln (W(\pi)/W(0)) &= -t+\delta\cos(2t)/2\big|_0^\pi\\
W(\pi) &=e^{-\pi}
\end{align*}
\fi
Using  Floquet theory \cite{guckenheimer2013nonlinear}, the stability of the system (\ref{eq:LTV}) is determined
by the eigenvalues of $M$, $\lambda = (\mathrm{tr}(M)\pm\sqrt{(\mathrm{tr}(M)^2-4e^{-\pi})})/2$ where the instability results if either eigenvalue has a modulus greater than one (\fref{floquet}).
Since the closed-form solution of the eigenvalues was difficult to obtain, we turned to numerical simulation. We determined the stability range for $\delta\leq \delta^*(\approx 3.2)$ and verified that the product of the eigenvalues is indeed $e^{-\pi}$ for all $\delta$.

%=================================================================
%=================================================================

\begin{remark}
  Note that the maximum value of $\delta$ ensuring stability based on the Floquet analysis is $\delta^* \approx 3.2$, which is considerably higher than the bound we obtained through Lyapunov analysis in the previous section ($\delta^\dagger\approx 0.43$). This ``daylight'' between our analytical and numerical analyses arises from the conservative nature of the Lyapunov function approach. Instead of requiring that $\dot V<0$ $\forall t$ one must only ensure that over any period, $V$ decreases, namely $V(t+\pi)<V(t)$, $\forall t$. We leave finding a tighter theoretical bound to future work, as the direct approach of explicitly integrating the flow of \eqref{eq:LTV} appears nontrivial.
\end{remark}

\begin{remark}
  As noted in Remark \ref{exp_stable}, the system \eqref{eq:LTV} is exponentially stable for sufficiently small $\delta$.  Our numerical simulations provide more insight into performance, e.g., allowing us to select the active sensing, $a$, and feedback gain, $k$ to maximize the convergence rate by ensuring  $\delta\in[\delta_1, \delta_2]$; see \fref{floquet}.
  
\end{remark}

Linear analysis only provides insight into the local behavior around the limit cycle. To understand the global behavior of the nonlinear system (\ref{eq:forced_simple_system}), in the next section, we adopt a numerical approach to determine the domain of attraction (DoA), as in the general nonlinear case, the DoA does not admit an analytical representation.
Additionally, techniques \cite{kang2023data,vannelli1985maximal} developed for autonomous systems are in general not trivial to extend to nonautonomous systems.

%=========================================================
\subsection{Numerical estimate of  domain of attraction}
\begin{figure}[htbp!]
\centering
\includegraphics[width=\columnwidth]{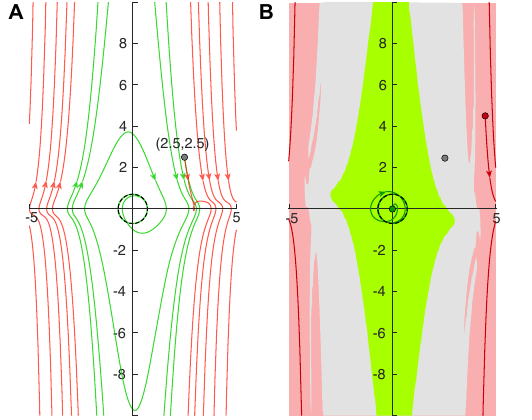}
\caption{Domain of attraction (DoA) for the nonlinear system (\ref{eq:LTV}) for $\delta = 1/2$ with $k =1, a= 1/\sqrt{2}$. (A) Trajectories from the same initial location $(x_0,z_0)=(2.5,2.5)$ (gray marker) initiated at $t=0$ (green) converges whereas the one initiated at $t=7\pi/8$ (red) diverges from the periodic solution, $\xnonlin^*(t)$ (black dashed circle).
(B) The green region is the conservative DoA, $D^*$, from within which all initial conditions converge to the periodic solution irrespective of the $t_0$. 
The red region denotes the set of all initial conditions that diverge irrespective of $t_0$. The convergence (or not) of trajectories whose initial conditions lie within the gray region depends on the initial time as illustrated in (A).
} 
\label{doa}
\end{figure}

In general, for a nonautonomous system, the convergence of state trajectories depends on both the initial conditions and the initial time.
%For some initial conditions, convergence may depend on the initial time.
\bd{def:DoA} Let $\xnonlin^*(t)$ be the periodic solution and, $\varphi_{\xnonlin_0,t_0}(t)$ the solution of the nonautonomous, nonlinear system (\ref{eq:forced_simple_system}) with the initial condition $\xnonlin_0(t_0)$. Given the system (\ref{eq:forced_simple_system}) is locally asymptotically stable with respect to $\xnonlin^*$, the domain of attraction (DoA) \cite{mct} of $\xnonlin^*$ for a given initial time $t_0$ is given by the set:
\begin{equation*}
    D(t_0):=\{\xnonlin_0\in\R^2 \mid \lim_{t \to \infty}\varphi_{\xnonlin_0,t_0}(t)-\xnonlin^*(t) =0\}.
\end{equation*}
\ed
The set of initial conditions for which the system converges irrespective of the initial time is defined as follows. 
\bd{def:abs_DoA} The conservative domain of attraction for the system (\ref{eq:forced_simple_system}) is given by the set:
\begin{equation*}
    D^* = D(\xnonlin^*):=\cap_{t_0\in[0,2\pi]}D(\xnonlin^*,t_0).    
\end{equation*}
\ed
Note that the original nonlinear system (\ref{eq:forced_simple_system}) is $2\pi$-periodic, and regardless of the initial time, $t_0\in[0,2\pi]$, trajectories with initial states, $\xnonlin_0$ within $D^*$  (green region in \fref{doa}) always converge to the periodic orbit
 $\xnonlin^*(t))$. Trajectories with initial states in the set $\{\R^2\setminus \cup_{t_0\in[0,2\pi]}D(\x^*,t_0)\}$ always diverge, whereas the convergence or divergence of the trajectories originating from the set $ \{\cup_{t_0\in[0,2\pi]}D(\x^*,t_0)\setminus D^*\}$ depends on $t_0$.

 %\njccom{Should the above be $\R^2$? in the def'n of $\overline{D^*}$?}

%Since we proved (Section \ref{sec:lyap}) the system (\ref{eq:forced_simple_system}) is locally asymptotically stable with respect to $\x^*$ 

%=================================================================
%=================================================================

\section{Discussion}
\label{sec:discussion}
%\dbcom{Please check the discussion}
 In this manuscript, we present a control strategy inspired by the active sensing movements observed in animals. While the system's structure is motivated by the locomotion dynamics of weakly electric fish, the framework can be adapted to model the behaviors of other animals with translationally invariant plant dynamics and appropriately modeled output measurements.

 Our approach extends beyond the realm of biology to a broader class of control problems dealing with lack of observability. For example, Brivadis \textit{et al.} \cite{brivadis2021new} addressed challenges related to the lack of observability at a target point (origin) arising from an output nonlinearity. In their system, observability was possible except at the origin and the system matrix, $A$ was assumed to be invertible. 
 In contrast, our system is unobservable along the entire x-axis and no restriction of invertibility on $A$ was imposed (in fact $A$ is not invertible). Instead of stabilizing to the origin (which we proved is impossible using output feedback for our system), we designed the input to play a dual role---ensuring both observability and output stabilizability to a periodic orbit around the origin. With suitable parameter tuning, the periodic orbit can be made arbitrarily small (albeit with a reduced convergence rate), effectively ensuring that the trajectories remain within an arbitrarily close neighborhood of the origin.

In conventional control engineering, the design of output feedback controllers often relies on the separation principle, which allows for the independent design of observers (based on sensor inputs) and controllers (assuming full-state measurements). However, this principle often fails to apply to general nonlinear systems. Our approach offers an alternative to this method. Conceptually, ``active sensing” is the opposite approach to applying a separation principle: control inputs are specifically designed to excite sensors, effectively enhancing the information gleaned from sensors and thereby improving feedback control. %In this way, the control inputs serve a dual purpose, actively engaging the sensory system to preserve observability while guiding the system toward stabilization.

For observable nonlinear systems, it is well-known that generic feed-forward (open-loop) inputs are sufficient to guarantee observability~\cite{2007uniform}. In that vein, we address the lack of observability in our system \eqref{eq:simple_system}
with a continuous open-loop active sensing input (coupled with an output feedback term for stability).
This choice reflects how animals engage in continuous active sensing.
However, depending on sensory salience, animals often opt to perform these sensing movements intermittently, rather than continuously \cite{biswas2023mode}. 
Moreover,
we did not impose any constraint on the energy budget, but animals likely tailor their movements for economy.
%The current model does not account for such intermittent behaviors, nor does it impose constraints on energy budgets, which animals must manage during task execution. 
%In the future, it would be worth exploring an optimal strategy of intermittent control that accounts for uncertainty in sensor measurements.
Thus in future work we aim to explore optimal robust strategies for intermittent sensing that balance energy efficiency with sensory needs, taking into account uncertainty in sensor measurements.

While recent works \cite{cellini2023empirical, davis2023influence, chen2020tuning}, including our own \cite{biswas2023mode}, have proposed heuristic models of active sensing behaviors, the present study establishes a more theoretically grounded approach to integrated control and sensing. We hope this foundation offers new insights into biologically plausible control strategies and inspires alternative engineering designs that could more effectively integrate sensing and control.

\section*{Acknowledgement}

%This work was supported by the Office of Naval Research under the grant N00014-21-1-2431 awarded to NJC and EDS. EDS also acknowledges support from grant AFOSR FA9550-21-1-0289.

This work was supported by the Office of Naval Research under grant N00014-21-1-2431 awarded to NJC and EDS, the Air Force Office of Scientific Research under grant FA9550-21-1-0289 to EDS, and the National Institutes of Health under grant U01NS131438 to NJC.
Special thanks to
Andrew Lamperski for useful discussions on proving stability.

\IEEEtriggeratref{25}

%\bibliographystyle{ieeetr}
%\bibliography{observability_cascade}

\end{document}